\documentclass[review]{elsarticle}

\journal{Journal of \LaTeX\ Templates}

\usepackage[inkscapeformat=png]{svg}

\usepackage{amsfonts}
\usepackage{amsmath}
\usepackage{amsmath} 
\usepackage{amssymb} 
\usepackage{algorithm}
\usepackage{algpseudocode}
\usepackage{setspace}
\usepackage{fullpage}
\usepackage{microtype}

\begin{document}

\begin{frontmatter}

\title{TagGAN: A Generative Model for Data Tagging}

\author[add1,add3]{Muhammad Nawaz}

\author[add2]{Basma Nasir}

\author[add2,add3]{Tehseen Zia} 
\author[add4]{Zawar Hussain}
\author[add1]{Catarina Moreira}
\ead{Catarina.PintoMoreira@uts.edu.au}

\address[add1]{Data Science Institute, University of Technology Sydney, Australia}
\address[add2]{COMSATS University Islamabad, Pakistan}
\address[add3]{Medical Imaging and Diagnostics Lab, National Center of Artificial Intelligence, Pakistan}
\address[add4]{Macquarie University, Sydney, Australia}

\begin{abstract}
Precise identification and localization of disease-specific features at the pixel-level are particularly important for early diagnosis, disease progression monitoring, and effective treatment in medical image analysis. However, conventional diagnostic AI systems lack decision transparency and cannot operate well in environments where there is a lack of pixel-level annotations. In this study, we propose a novel Generative Adversarial Networks (GANs)-based framework, TagGAN, which is tailored for weakly-supervised fine-grained disease map generation from purely image-level labeled data. TagGAN generates a pixel-level disease map during domain translation from an abnormal image to a normal representation. Later, this map is subtracted from the input abnormal image to convert it into its normal counterpart while preserving all the critical anatomical details. Our method is first to generate fine-grained disease maps to visualize disease lesions in a weekly supervised setting without requiring pixel-level annotations. This development enhances the interpretability of diagnostic AI by providing precise visualizations of disease-specific regions. It also introduces automated binary mask generation to assist radiologists. Empirical evaluations carried out on the benchmark datasets, CheXpert, TBX11K, and COVID-19, demonstrate the capability of TagGAN to outperform current top models in accurately identifying disease-specific pixels. This outcome highlights the capability of the proposed model to tag medical images, significantly reducing the workload for radiologists by eliminating the need for binary masks during training.
\end{abstract}


\end{frontmatter}

\section{Introduction}
{
Pulmonary diseases, including Tuberculosis (TB), pneumonia, and COVID-19, remain a critical global health challenge. Together, these conditions claim millions of lives annually, underscoring the need for early and accurate diagnosis to reduce mortality rates \cite{1,2}.
Chest radiographs are the gold standard for diagnosing pulmonary conditions, as radiologists use them to detect and localize abnormalities \cite{3}. However, the shortage of qualified specialists, combined with heavy workloads exacerbated by the COVID-19 pandemic, limits regular monitoring of disease progression \cite{4,5}. To address this, Artificial Intelligence (AI) systems have been developed in clinical practice to streamline and accelerate the diagnostic process, even assisting less-experienced practitioners \cite{6}. These methods excel at identifying and classifying diseases. However, the decision-making process behind these algorithms remains opaque to both programmers and medical specialists \cite{7,8}.  A major concern when deploying classical algorithms in critical applications is their 'black box' nature, which undermines trust \cite{9}.

To address this issue, explainable artificial intelligence (XAI) aims to limitations of the black-box AI models by improving decision transparency \cite{10}. In this context,  Visualization of the attributes (VA) focused on efficiently detecting and displaying disease-related features in medical images during classification. In recent years, this field has gained significant attention due to its wide range of applications, including weakly supervised segmentation \cite{11}, explainable artificial intelligence \cite{12}, and the identification and visualization of disease-specific features in medical imaging \cite{13}. However, despite these advances, methods that rely on neural network classifiers often deliver suboptimal results. They tend to focus on a limited set of image features, which can cause them to miss important regions of interest \cite{14}.  While current Visualization of Attributes techniques, particularly those using GANs, have shown promise, they are still limited in scope and often struggle to accurately localize multiple disease-specific lesions across various parts of an image \cite{15,16}. These limitations inherent in both classification and generative algorithms highlight the need for a model capable of generating detailed disease maps that effectively and comprehensively identify all disease features across arbitrary regions of an image. While pixel-level annotation is critical for creating accurate and interpretable diagnostic algorithms, most medical imaging databases offer only image-level data, with some providing bounding box annotations at best. This lack of detailed annotations poses a significant challenge, particularly for diseases like COVID-19, which require continuous monitoring at pixel-level precision. Traditional radiological techniques are insufficient for such fine-grained analysis, highlighting the need for robust AI systems capable of real-time, pixel-level disease progression monitoring \cite{17,18}.The goal of fine-grained X-ray analysis is to detect all disease-specific features at the pixel level, regardless of their location within an image. A key challenge in developing such algorithms is the lack of sufficient pixel-level annotation data in medical imaging. Most databases provide either images or at most, bounding box-level annotations. Manually labeling these images for more detailed annotations is both labor-intensive and costly, further hindering the development of accurate and automated diagnostic tools. 

Recent works proposed interpretable GAN-based methods for identifying and visualizing domain features by incorporating disease maps into input radiographs \cite{15,16}. However, these approaches often fail to preserve key anatomical structures during domain translation, resulting in inaccurate mappings of specific radiographs. This leads to the generation of false-positive pixels, which appear as random noise in disease maps and are mistakenly interpreted as real disease features, potentially causing diagnostic errors. In summary, current methods face significant limitations, including: (a) prioritizing highly discriminative disease features while excluding finer, less discriminative details; (b) inefficiencies in decision interpretability; and (c) an inability to train in weakly supervised settings where binary annotations are absent. Consequently, there is a need for an explainable data-tagging technique that can be trained in a weakly supervised manner, without pixel-level labels, while still producing pixel-level annotations of disease features.  In this study, we introduce TagGAN, a generative network designed to address the challenges of visualizing disease-specific features and automatically generating pixel-level labels without relying on binary masks for training. To the best of our knowledge, TagGAN is the first algorithm capable of annotating or tagging medical images at the pixel level without the use of binary masks, representing a significant advancement in medical imaging technology. In this article, the terms "annotation" and "tagging" are used interchangeably.The main contributions of TagGAN are the following:

\begin{itemize}
\item \textbf{Weakly supervised learning:} TagGAN is capable of being trained in a weakly supervised setting, utilizing only image-level labels. Despite the absence of pixel-level annotations, it accurately detects and localizes class-specific pixels at a fine-grained level, enabling the generation of pixel-level labels for X-ray inputs.
\item \textbf{Model interpretability:} The model offers a high degree of interpretability, as the underlying mechanisms responsible for true-positive pixel identification, feature visualization, and data tagging are transparent and can be easily understood, even by non-experts.
\item \textbf{Multi-domain generalization:} TagGAN can be trained across multiple domains, allowing it to learn multi-class mappings and function as a domain translation algorithm for different classes of medical data.
\end{itemize}

The remainder of this article is organized as follows. Section 2 presents the most relevant existing research. Section 3 provides the practical implementation of the proposed methodology and architecture while Section 4 explains the implementation, which includes the baseline model, the training of the algorithm, and the datasets used in the experiment. Section 5 discusses the quantitative and qualitative findings and how they compare to one another, and Section 6 concludes this work.

}
\section{Related Work}
{Classification of diseases by medical image analysis is a vital area of research in the healthcare sector. Imaging analysis aided by AI can detect COVID-19 and TB quickly, lessening the effect of these diseases. The foregoing discussion of the features of X-ray images forms the basis of an AI-based method described here for accurately interpreting TB and COVID-19 cases from the input CXR radiograph with the aid of professional radiologists. Several experiments have been conducted to lessen the burden on doctors and radiologists for analysis and data tagging by using artificial intelligence and machine learning for faster COVID-19 and earlier TB screening and tagging. This section summarizes existing disease detection and feature visualization methods, including both discriminative and generative approaches.}

\subsection{Generative Adversarial Networks (GANs)}
{
GANs are two-part designs, with the generator network programmed to generate new content and the discriminator network distinguishing the data coming from both real and false data distribution \cite{19}. Generator and discriminator models are built concurrently, with the generator aiming to fool the discriminator by simulating genuine occurrences. In numerous fields, GANs have proven to be highly effective \cite{20,21}. Such examples include face image synthesis and domain interpretation. While existing GAN-based disease feature visualization methods use image-to-image translation to move data from one domain to another, they require labeling information that is not available in the raw data. Consequentially, these methods are often manual and incompetent for solving various problems, such as describing the decision-making process and generating pixel-level annotations. To facilitate domain translation, researchers have introduced the Cycle consistent GAN (CycleGAN), the first unpaired image-to-image translation model that converts abnormal to normal classes without requiring paired normal images during training \cite{22}. Our proposed method is fundamentally based on this approach. GANs have been used for a variety of applications, especially in healthcare and medical imaging analysis. Disease identification, organ segmentation (such as the heart or liver), artery network segmentation, and disease impact visualization, in the context of XAI, are just a few examples \cite{23,24}.
}

\subsection{Disease Attributes Visualization}
{
Visual attribution refers to the process of locating and representing in an image the specific signs of disease that help define the category of an input image. Deep features from classifiers are commonly used for segmentation or localization in tasks with limited supervision \cite{25}. Class Activation Mapping (CAM) \cite{26} is a popular technique with similar goals; it generates a class-specific feature map by incorporating a pooling layer with a global average into a system to reduce the number of networks that collectively make up the feature set. Because the results for each value are independently calculated, a CAM model cannot be used to efficiently generate a concise map for attribute visualization. Using these methods, it is difficult to pinpoint the pixel-level characteristics of a certain disease. Using a back propagation approach to restore gradients to their original locations in the input CXR, saliency maps are a unique method for identifying and visualizing illness attributes.

However, they are dependent on classification, thus they do not give a whole picture of the region of interest, and they also overlook some key aspects of diseases. Since the approach is concerned with the architecture's ultimate feature space, it necessitates pre-processing of the network prediction. As researchers in an article \cite{27} point out that the saliency maps only represent the places that networks stare at often while making a certain decision. Another methodology, Grad-CAM \cite{28}, only depicts high-density areas where the system focuses its attention while creating illness maps. Nothing in the image should be taken as a suggestion that the location portrayed has any infectious elements. If the data is missing critical information or there are no binary masks to use to train the algorithm, these discriminative-based approaches cannot operate in a weakly supervised setting.

To solve these issues with discriminative models, researchers developed a generative-based explainable technique called Visual Attribution GAN (VA-GAN) \cite{15}. The VA-GAN uses the concept of visual interpretation to pinpoint disease features, but this has unintended results because the mapping function translates every abnormal image into any image of a normal class. The resulting normal CXR has noises that are not a pair of input diseased CXRs. The discriminator of VA-GAN, however, labels the newly generated CXR as normal, even though it does not match either of the input aberrant images. Because of this, medical data will never be tagged by a model that cannot detect disease at the pixel-level and can not preserve anatomical features. In reality, this structure is founded on Wasserstein GAN \cite{29}, which does not reliably maintain the originality of input images because of only the forward phase. Importantly, this model cannot build binary masks because it only takes into account a small subset of the properties of the domain being shown. Another generative-based disease visualization method, ANT-GAN \cite{16}, was proposed as a means of dealing with the aforementioned problems and producing normal images while retaining the contents of the input abnormal images. This approach makes an effort to account for all attributes unique to the domain, resulting in an abnormal pair of input data. Nevertheless, because it is based on the disease map addition concept, this approach is limited in how it may be used to explain medical data because it does not account for disease identification at the pixel level. To create binary masks and train algorithms for interpretability, paired data is required, but this is not always available. Particularly in radiology, where it would be impractical to track down a patient's healthy and abnormal chest X-rays, this is an issue. 

As a result, an explainable model that can identify the illness traits and automatically tag the medical data is required in the setting of weakly supervised learning to detect the disease at a fine-grained level. To find disease-oriented features at a fine-grained level, show the pixel-level information of those locations in the input, and tag the input CXR radiograph, the framework should be trained on CXR with domain-level labels without binary masks. Therefore, our proposed TagGAN did not need paired data to learn modeling among anomalous and normal distribution, identify disease consequences at pixel-level, depict abnormalities of the disease, and provide annotation of input CXR radiograph, unlike the CycleGAN, VA-GAN, and ANT-GAN models. Recent works in both generative and discriminative models are summarized in Table \ref{tab:example1}, along with details regarding the challenges at hand, the methods chosen, and the limitations of these methods.

\begin{table}[h!]
\centering
\caption{\label{tab:example1}Summary of Methods}
\bigskip
\begin{tabular}{| p{2cm}|p{4cm}|p{2cm}|p{4cm}|}
\hline
\textbf{Reference} & \textbf{Methodology} & \textbf{Algorithm} & \textbf{Remarks} \\
\hline
Baumgartner, C. F., et al. \cite{15} & To show the effects of the disease without resorting to classification, this model generates disease maps from diseased CXR. It combines the input image with the created change map to get the normal distribution. & WGAN & (1) Causes false positive results, (2) adds random noise in the visualization map, and (3) cannot tag data \\
\hline
Sun, L., et al. \cite{16} & A standard CycleGAN-based method for identifying and visualizing lesions, wherein abnormal data is complemented with a disease map to provide a computed tomography (CXR) image of the lesion in its nonabnormal form. & CycleGAN & (1) Limited interpretability, (2) unable to identify disease at pixel-level, and (3) cannot be used to tag data  \\
\hline
B. Zhou, et al. \cite{26} & This method accomplishes interpretability by first shrinking the network's feature set and then averaging their data with a pooling layer. To this goal, class-specific activation maps are produced. & CAM & (1) The network's prediction requires post-processing, and (2) cannot be used to tag data\\
\hline
\end{tabular}

\end{table}
}

\section{Methodology}
\subsection{TagGAN}
{
Our goal is to develop an automated pixel-level annotation model for healthcare, reducing the need for labor-intensive and costly manual tagging. To label diseased CXRs accurately, our algorithm generates a disease-specific feature map by transforming a diseased image into its healthy counterpart. Figure \ref{fig:Picture1} illustrates the architecture of TagGAN, which leverages domain translation to identify, visualize, and tag disease-oriented features in CXR images at the pixel level in a semi-supervised setting, without requiring binary masks.
Previous methods attempting to achieve similar goals have struggled due to their reliance on paired normal chest radiographs for training, which are often unavailable. They typically superimposed disease maps onto source images, resulting in noise and failing to produce paired normal radiographs with pixel-level disease information. Our approach, TagGAN, addresses these limitations by classifying data into two groups: diseased CXRs as input and healthy CXRs as baselines. For instance, TagGAN can detect multiple regions of interest (ROIs) in the chest radiographs of TB patients where disease features appear in distinct areas. The disease map is \(M(x_i, c)\) is generated by subtracting it from the input diseased image \(x_i\) \ref{eqn_1}:

\begin{equation}
\label{eqn_1}
    y_i = x_i - M(x_i, c)
\end{equation}

where $x_i$ represets a diseased sample, $`c`$ is domain label, $M(x_i, c)$ is the disease map, and $y_i$ is the corresponding baseline instance of input $x_i$ obtained by subtracting $M(x_i, c)$ from input $x_i$. However, due to the lack of paired medical image data, we implement a cyclic consistency process to map diseased X-rays to healthy images and back. Our model first localizes, pixel by pixel, all the diseased characteristics in the entire image, even if the disease impact is occurring in many locations, and then generates the labeled image from this compact disease map. TagGAN’s architecture comprises two cycles: the forward-cycle GAN (fGAN) and the backward-cycle GAN (bGAN). In fGAN, the generator $G_{\text{A-to-M}}$ receives a diseased input \(x_i\) with domain label \(c\) and produces a disease map \(M(x_i, c)\). The transformed instance \(y_i\) is then assessed by the discriminator $D_{\text{Normal}}$, which distinguishes between true and synthetic baselines. The adversarial loss for training $G_{\text{A-to-M}}$ is give by \ref{eqn_2}:

\begin{equation}
\label{eqn_2}
\begin{aligned}
     L_{fGAN}= \mathbb{E}_{y\sim p_{data(y)}}\left [ \log (D_{Y}(y_i)) \right ]\\+\mathbb{E}_{x\sim p_{data(x)}}\left [ \log (1- D_{Y}(x_i-M(x_i - c))) \right ]
     \end{aligned}
\end{equation}

The backward cycle in bGAN regenerates the diseased instance \(x_i\) from the healthy instance
\(y_i\), preserving the anatomical features of the original CXR. This iterative process ensures that the generated healthy image retains only the disease-specific features, ultimately achieving fine-grained disease mapping at the pixel level. The overall objective function for optimizing the model is given by \ref{eqn_3}.

\begin{equation}
\label{eqn_3}
\begin{aligned}
     L_{bGAN} = \mathcal{E}_{x\sim p_{data(x)}}\left[\log(D_{X}(x))\right]\\ +
     \mathcal{E}_{x\sim p_{data(x)}}\left[\log(1- D_{X}(G_{N\_to\_A}(y_{i} - 
     G_{A\_to\_M}(x_{i}, c)), b))\right]
     \end{aligned}
\end{equation}

We adopted the cycle consistency loss function, as presented in Equation \ref{eqn_4}, to learn the mapping of the input diseased CXR to its corresponding healthy counterpart and subsequently back to the original diseased image.

\begin{equation}
\label{eqn_4}
\begin{aligned}
     L_{cyc}(G_{A\_to\_M}, G_{N\_to\_A}) = \mathbb{E}_{x \sim p_{data(x)}} \left[ || G_{N\_to\_A}(x_{i} - G_{A\_to\_M}(x_{i}, c)) - x ||_{1} \right] \\+ \mathbb{E}_{y \sim p_{data(y)}}
     \left[ || G_{A\_to\_M}(G_{N\_to\_A}(y_{i}, b)) - y ||_{1} \right]
     \end{aligned}
\end{equation}

Finally, Equation \ref{eqn_5} shows the objective function with optimization of the generators and discriminator models.

\begin{equation}
\label{eqn_5}
\begin{aligned}
     L(G,D)=L_{fGAN} +L_{bGAN} + L_{cyc}
     \end{aligned}
\end{equation}

At convergence, TagGAN produces a disease map\(M(x_i, c)\) that highlights disease-specific features at a granular level. Finally, we apply a constraint function to transform the disease map into an annotated binary mask, as shown in Figure \ref{fig:Picture1}.

\begin{figure}[h!]
\centerline{\includegraphics[width=13cm,height=12cm]{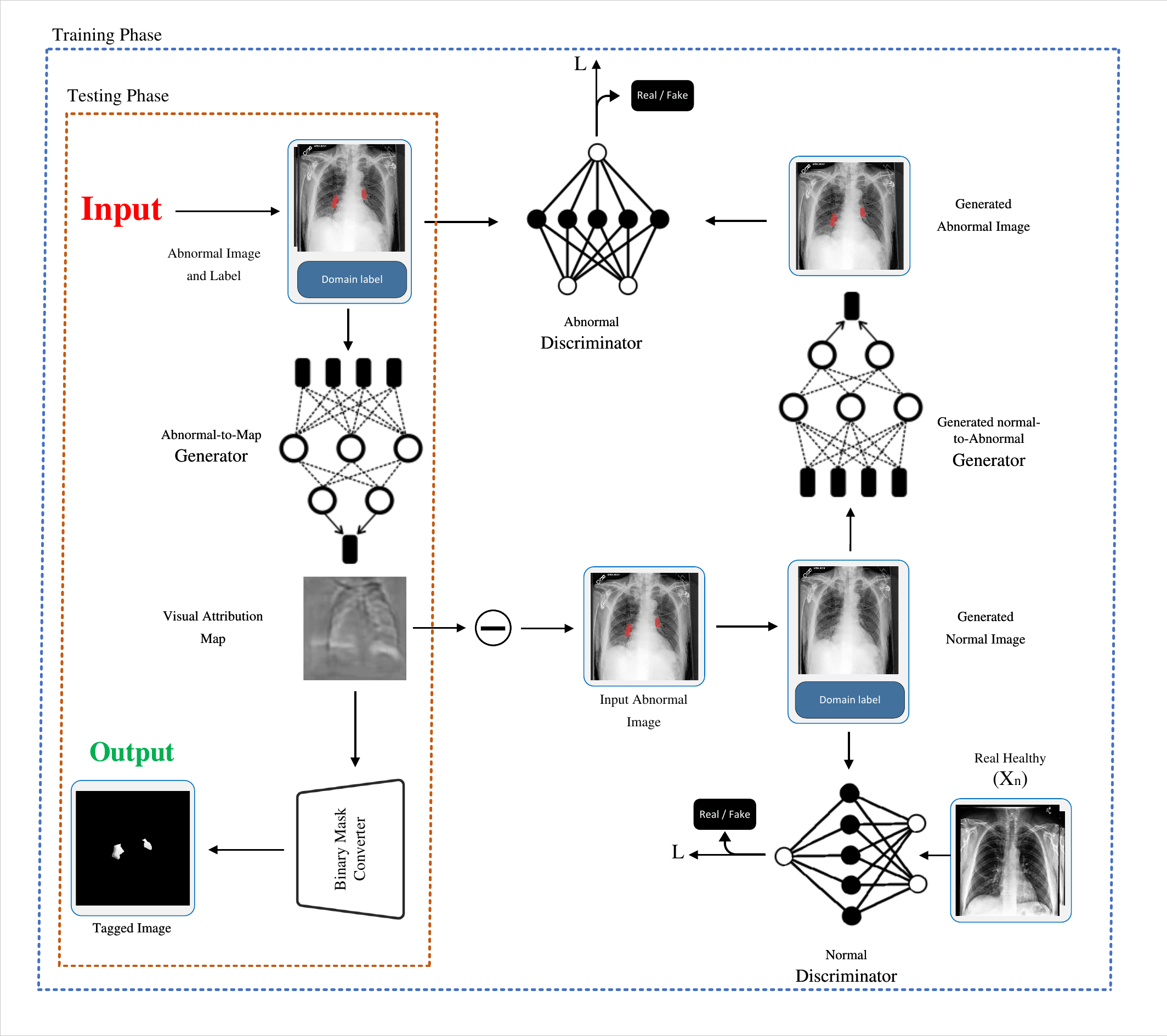}}
\caption{Architecture of TagGAN: (a) during the training phase, input is diseased CXR with an image-level label to an Abnormal-to-Map generator which generates a feature map, called visual attribution map, and when this map is subtracted from input abnormal image to covert it into a normal CXR pair. This generated normal pair is fed to Normal Discriminator which is trained on original healthy data. For backward cycle, this generated normal CXR is input to second generator model which converts back into abnormal and second discriminator, trained on original abnormal data, distinguishes it. (b) during testing phase, input is diseased CXR with an image-level label to Abnormal-to-Map generator which generates a feature map to visualise disease-specific features and a constraint function coverts this map into binary mask}
\label{fig:Picture1}
\end{figure}
}

\begin{algorithm}[h!]
\caption{: Number of steps $k$ is a hyperparameter, but in my case $k =$ number of real samples. And $c$ is the disease label of disease $l$, whereas $l$ is the number of diseases/domains.}
\begin{algorithmic}
\For{number of training iterations}
    \For{$k$ steps}
        \begin{itemize}
            \setlength{\itemsep}{0pt} 
            \item Sample minibatch of size $A$ $\{x_1, x_2, \ldots, x_a\}$ from image set $X$.
            \item Sample domain labels $c$ of size $l$ $\{c_1, c_2, \ldots, c_l\}$ from label set $X_l$.
            \item Sample minibatch of size $N$ $\{y_1, y_2, \ldots, y_n\}$ from image set $Y$.
            \item Generate domain-specific VA map $M$ and synthesize target images $\hat{y}$:
        \end{itemize}
        \vspace{-10pt}
        \begin{align*}
            M_{i=1}^l &= \{G(x_1, c_1), G(x_2, c_1), \ldots, G(x_a, c_l)\}, \\
            y_{j=1}^n &= \{M_1{-x_1}, M_2{-x_2}, \ldots, M_l{-x_a}\}.
        \end{align*}
        \begin{itemize}
            \setlength{\itemsep}{0pt}
            \item Reconstruct $X$ from generated images $y$ and domain label:
        \end{itemize}
        \vspace{-10pt}
        \begin{align*}
            X_{i=1}^n = \{F(y_1,c_1), F(y_2,c_1), \ldots, F(y_n,c_l)\}.
        \end{align*}
        \begin{itemize}
            \setlength{\itemsep}{0pt}
            \item Calculate domain classification loss on real and fake images:
        \end{itemize}
        \vspace{-10pt}
        \begin{align*}
            L_{dcl}^r &= E_{(x,c)} [-\log D(x|c)], \\
            L_{dcl}^f &= E_{(x,c)} [-\log D(G(x,c)|c)].
        \end{align*}
        \begin{itemize}
            \setlength{\itemsep}{0pt}
            \item Update $D_Y$ model:
        \end{itemize}
        \vspace{-10pt}
        \begin{align*}
            \nabla_{\theta_{DY}} \frac{1}{A} \sum_{i=1}^{A} \left[ \log D_Y(y_i) + \log(1 - D_Y(G(x_i,c_i) - x_i)) \right].
        \end{align*}
        \[
        \text{Such that } \theta \text{ represents network parameters, } D_Y(y_i) = \text{true, and } D_X(G(x_i, c_i)) = \text{false, for } i = 1, 2, \ldots, A.
        \]
        \begin{itemize}
            \setlength{\itemsep}{0pt}
            \item Update $D_X$ model:
        \end{itemize}
        \vspace{-10pt}
        \begin{align*}
            \nabla_{\theta_{DX}} \frac{1}{A} \sum_{i=1}^{A} \left[ \log D_X(x_i) + \log(1 - D_X(F(y_i,c_i))) + L_{dcl}^r \right].
        \end{align*}
        \[
        \text{Such that } \theta \text{ represents network parameters, } D_X(x_i) = \text{true, and } D_X(F(y_i, c_i)) = \text{false, for } i = 1, 2, \ldots, A.
        \]
        \begin{itemize}
            \setlength{\itemsep}{0pt}
            \item Update $G$ and $F$ models:
        \end{itemize}
        \vspace{-10pt}
        \begin{align*}
            \nabla_{\theta_G,\theta_F} \frac{\lambda}{A} \sum_{i=1}^{A} \left[ \left\| F(G(x_i,c_i) - x_i) - x_i \right\|_1 + \left\| G(F(y_i,c_i) - y_i) \right\|_1 + L_{dcl}^f \right].
        \end{align*}
        \[
        \text{Such that } \theta \text{ represents network parameters, } x_i \approx F(G(x_i, c_i)) - x_i \text{ and } y_i \approx G(F(y_i, c_i)) \text{ for } i = 1, 2, \ldots, A.
        \]
    \EndFor
\EndFor
\end{algorithmic}
\end{algorithm}

\section{Experimental Setup}
{
This section discusses the details of datasets, including CheXpert, TBX11K, and COVID-19, highlighting their characteristics and relevancy to underconsider problems. Also, outline the details about the baseline model and the training settings utilized to achieve optimal performance. Because the proposed model needs the GPU and a lot of processing power during training, we use the publicly available Google Colaboratory system from Google Research. The NVIDIA Tesla V100 GPU, 358.27 GB of storage space, and a minimum of 12 GB of RAM, expandable to 25 GB, were all features made available by Google Research.

\subsection{Datasets}
\subsubsection{CheXpert}
{
The Chest eXpert (CheXpert) \cite{30} dataset stands as a substantial and publicly available collection of multi-class chest X-rays. It comprises an extensive compilation of 224,316 chest radiographs drawn from 65,240 distinct disease patients. This dataset is particularly notable for its annotations, with each chest radiograph bearing class-level labels pertaining to the presence of 14 different categories. These categories encompass a range of conditions, including 13 various diseases alongside a normal class. The labeling scheme assigns values of 1 to indicate the presence of disease, 0 for uncertain instances, and -1 for the absence of disease. A pivotal step in our experimentation process involves reclassifying uncertain labels as indicative of the absence of disease. Moreover, the dataset encompasses chest radiographs sized at 390 x 320 pixels, accommodating both male and female subjects and capturing frontal and lateral views. In the context of our experiment, we have selectively included male chest X-rays with frontal views, resizing them to dimensions of 256 x 256 pixels.
}

\subsubsection{TBX11K}
{
 The TB X-ray dataset \cite{31} comprises chest radiographs of dimensions 512x512 pixels. Each X-ray image within this dataset is accompanied by a corresponding label that delineates whether the patient is deemed healthy, afflicted by a medical condition, or infected with tuberculosis. Notably, this dataset stands as a notable advancement in terms of being more contemporary, larger, and more extensively annotated in comparison to its predecessors, encompassing a total of 11,200 CXR images. The dataset division is structured as follows: The training set encompasses a total of 6,889 CXRs, the validation set comprises 2,087 CXRs, and the test set consists of 3,302 CXRs. It is important to note that while bounding box labels, often referred to as ground truth labels, are available for both training and validation sets, they are not provided for the test set. Worth clarifying is that the bounding box labels were not utilized for training purposes; rather, they were employed to ensure the efficacy of our proposed model.
}

\subsubsection{COVID-19}
{
 The COVID-19 dataset \cite{32} includes 900 chest radiographs from patients diagnosed with COVID-19 pneumonia. All images were resized to 512x512 pixels for consistency. Additionally, 4,600 radiographs depicting non-COVID pneumonia cases and 5,000 normal cases were included. The Tuberculosis and COVID-19 datasets are single-class datasets focused on their respective diseases, while CheXpert is a multi-class dataset. A few samples from all the datasets are shown in Figure \ref{fig:Picture2}.

\begin{figure}[h!]
\centerline{\includegraphics[width=10cm,height=11cm]{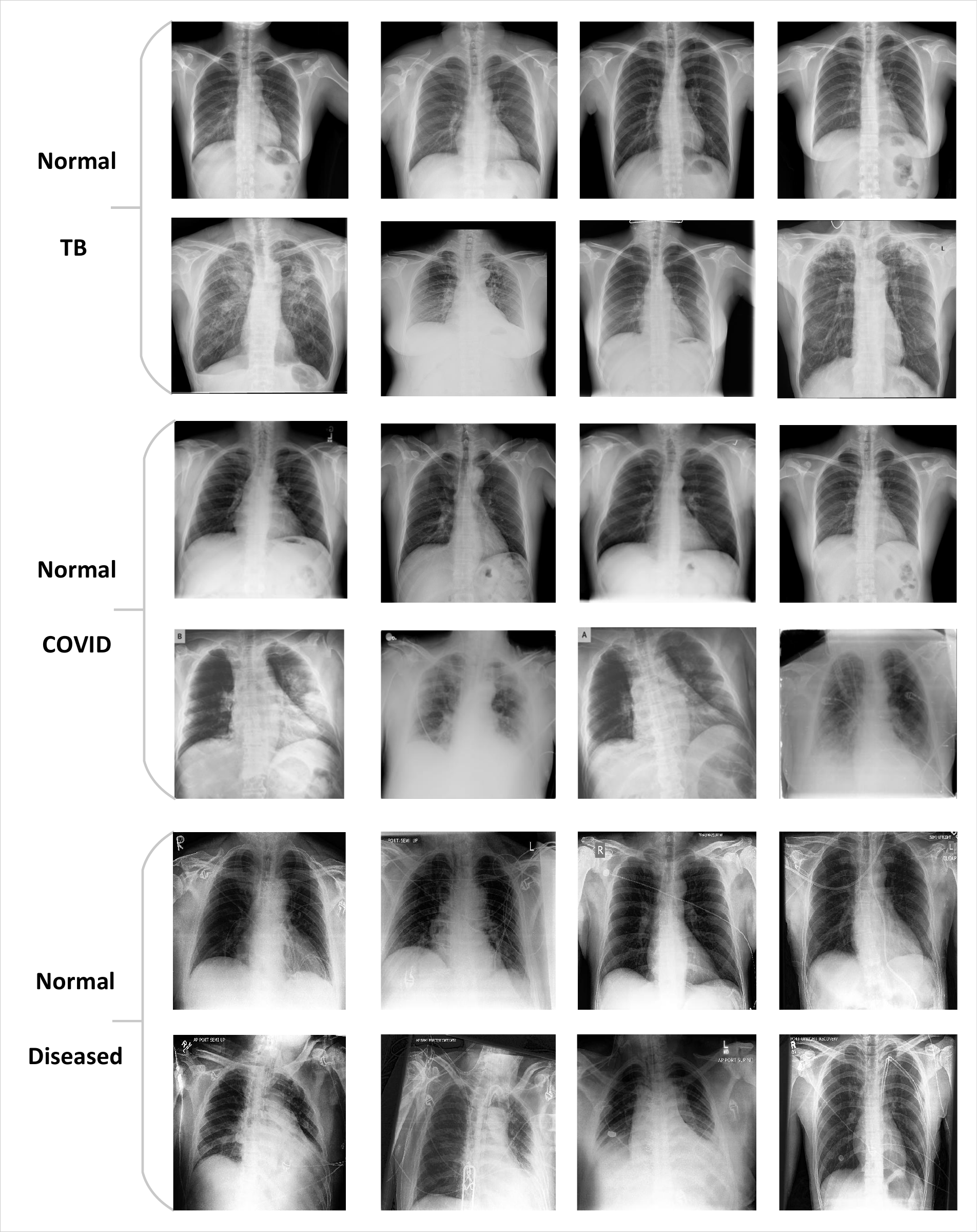}}
\caption{ Images from datasets}
\label{fig:Picture2}
\end{figure}
}

 \subsection{Baseline Model}
We use CycleGAN \cite{15} as a baseline model, an unpaired image-to-image translation framework capable of mapping between two domains. CycleGAN uses adversarial and cycle consistency losses to preserve input data integrity while learning relationships between two domains. We modified the original CycleGAN architecture to incorporate data-tagging capabilities, enabling pixel-level tagging and improving interpretability.

\subsection{Training Configuration }
{
 During training, the generator and discriminator models' learning weights are updated in an alternating fashion. To do this, we employ the ADAM optimizer. The batch size is set to 1 because research has shown that this is the optimal amount of data to feed into a GANs model at each stage but set learning rate to 0.001. The halting condition is reached after 10 validation epochs, during which time the accuracy of creating a disease map and a baseline normal image are validated. Loss for the generator and discriminator models are calculated using validation image input after five iterations. Forward cycle losses in both the abnormal-to-map generator and normal discriminator models are discussed. The number of training epochs was set at 100, and while the losses of generator and discriminator models fluctuated in the first 25 epochs, convergence set in after that. After finishing training, the discriminator model's loss had settled to around 0.5 on both the true distribution and the false data.}

\section {Experimental Results}
{
In this section, we present the results of our proposed model. We apply our model to COVID-19 and TB datasets and discuss the results from various aspects.  
 
To evaluate the proposed model and benchmark it against state-of-the-art methods, we use Intersection over Union (IoU) metric which is the standard for segmentation evaluation. IoU requiring pixel-level ground truth, which is unavailable in our setting, instead our model generates pixel-level labels from input images annotated only with bounding boxes. To address this, we use another metric, Percentage of Intersection (PoI), a novel metric quantifying the overlap between generated masks and bounding box annotations.

\subsection{Quantifying True Positive Detection}
Our main results, Table \ref{tab:truePositive}, shows performance of TagGAN in generating binary masks using only image-level annotations. Due to complexity of problem, designing a robust evaluation setting posed challenges. Especially, this comparison involves generated binary masks against bounding box ground truth, rather than simple binary mask vs. binary mask or bounding box vs. bounding box. In particular, bounding boxes in ground truth act as coarse boundaries encapsulating disease-specific regions, while the generated binary masks identify diseased-pixels across multiple localized areas inside those boundaries. To bridge this gap, we outlined bounding boxes around each pixel-level lesion and computed results under two settings: one with an equal number of bounding boxes as the ground truth and another using unequal setup. IoU scores were low, as expected, because calculation involves bounding boxes compared against pixel-level binary masks. These scores reinforces the nature of IoU as a metric, where large bounding boxes covering regions beyond diseased pixels naturally result in lower scores when calculated with precise pixel-level annotations. Rather than a limitation, this underscores complexity of problem and highlights the need for alternative metrics better suited to similar scenarios. PoI is metric which addresses this limitation by effectively capturing the percentage of overlap between ground truth and generated binary masks. Using PoI, our model achieved 80.06\% accuracy in identifying true positive.

 \begin{table}[!ht]
\centering
\begin{tabular}{|l|l|l|}
\hline
\textbf{TBX11K} & \textbf{PoI} & \textbf{IoU (\%)} \\
\hline
With equal number of bounding boxes & 70.78 & 27.76 \\
\hline
With an unequal number of bounding boxes & 80.06 & 0.12 \\
\hline
\end{tabular}
\caption{\label{tab:truePositive}Quantitative Results}
\end{table}

\subsection{Benchmarking Against Existing Methods}

Achieving high IoU scores when comparing binary mask to bounding box annotations is challenging due to fundamental difference in level of detail they represent. Consequently, IoU scores from pixel-level predictions are lower when evaluated against such ground truth, as the model’s outputs precisely localize disease-specific regions.
In current literature, a few algorithms can be used to address under consider problem. Grad-CAM and VA-GAN are two different nature methods but can be used to tackle this challenge. Grad-CAM achieves relatively high IoU scores, as it primarily select algorithm's focused coarse feature localization, aligning well with bounding box annotations. Similarly, VA-GAN provides binary mask outputs, though its localization capability is less fine-grained and noisy.
The core challenge, and key "puzzle" of this problem, lies in avoiding the use of binary masks during training while still achieving pixel-level identification of disease-specific features. This makes the problem challenging, yet our algorithm successfully addresses it, marking a new direction and novel contribution to the research community. To the best of our knowledge, TagGAN is the first algorithm to achieve a class-specific pixel identification accuracy of 80.06\% without requiring training on binary masks. Our work outperforms existing state-of-the-art methods by approximately 8\%, setting a new benchmark \ref{tab:example3}.

\begin{table}[!ht]
\centering
\begin{tabular}{|l|l|l|}
\hline

\textbf{TBX11K} & \textbf{PoI} & \textbf{IoU (\%)} \\
\hline
Grad-CAM \cite{28} & 52.15 & 41.62  \\
\hline
VA-GAN \cite{15}& 72.23 & 37.52 \\
\hline
TagGAN & 82.53 & 27.76 \\
\hline
\end{tabular}
\caption{\label{tab:example3}Quantitative Results Comparison}
\end{table}
}
 \subsection{Qualitative Results on Tuberculosis}
The tuberculosis dataset was used to qualitatively evaluate the algorithm, where chest X-rays, including tuberculosis-affected and normal cases, were input to the TagGAN algorithm to identify and visualize disease-specific pixels. Algorithm-generated disease maps serve as the foundation for assessing the model’s performance in weakly supervised disease detection.

The first column in Figure \ref{fig:Picture3} shows inputs X-rays, three TB samples and one normal. The second column represents the available bounding boxes as ground truth to indicate diseased areas, providing a standard for evaluating the model’s outputs.
The disease maps in the third column highlight the diseased pixels by isolating pathological features and subduing normal areas. During training, the algorithm transformed input diseased radiographs into normal counter pairs by subtracting the disease maps to learn disease patterns. During testing, the disease maps precisely localized only diseased regions and ignored normal pixels, as evidenced by blank disease maps for the normal X-ray input.
The tagged images in the fourth column represent the binary masks of the input images generated from the disease maps. These masks are a direct visualization of the algorithm’s capability to achieve the desired output as these are pixel-level identification of disease-specific features.
By highlighting affected regions in white and healthy areas in black, these generated binary masks precisely capture the diseased-oriented features within the input radiographs. These results validate the effectiveness of the model in achieving its core objective of generating binary masks directly from input images without relying on pixel-level annotations during training. The last two columns are visual representation of comparing the generated binary masks with ground truth annotations. The fifth column displays bounding boxes drawn on generated binary masks to localize the diseased regions. These bounding boxes encapsulate the pixels highlighted in the binary masks, providing a structured visualization of our model and used for comparison. The last column overlays the corresponding bounding boxes on the ground truth annotations, enabling a direct visual comparison. This side-by-side comparison highlights the alignment between the model's pixel-level predictions and the available bounding box ground truth.

 \begin{figure}[h!]
		\centering
		\includegraphics[width=1\linewidth, height=11cm]{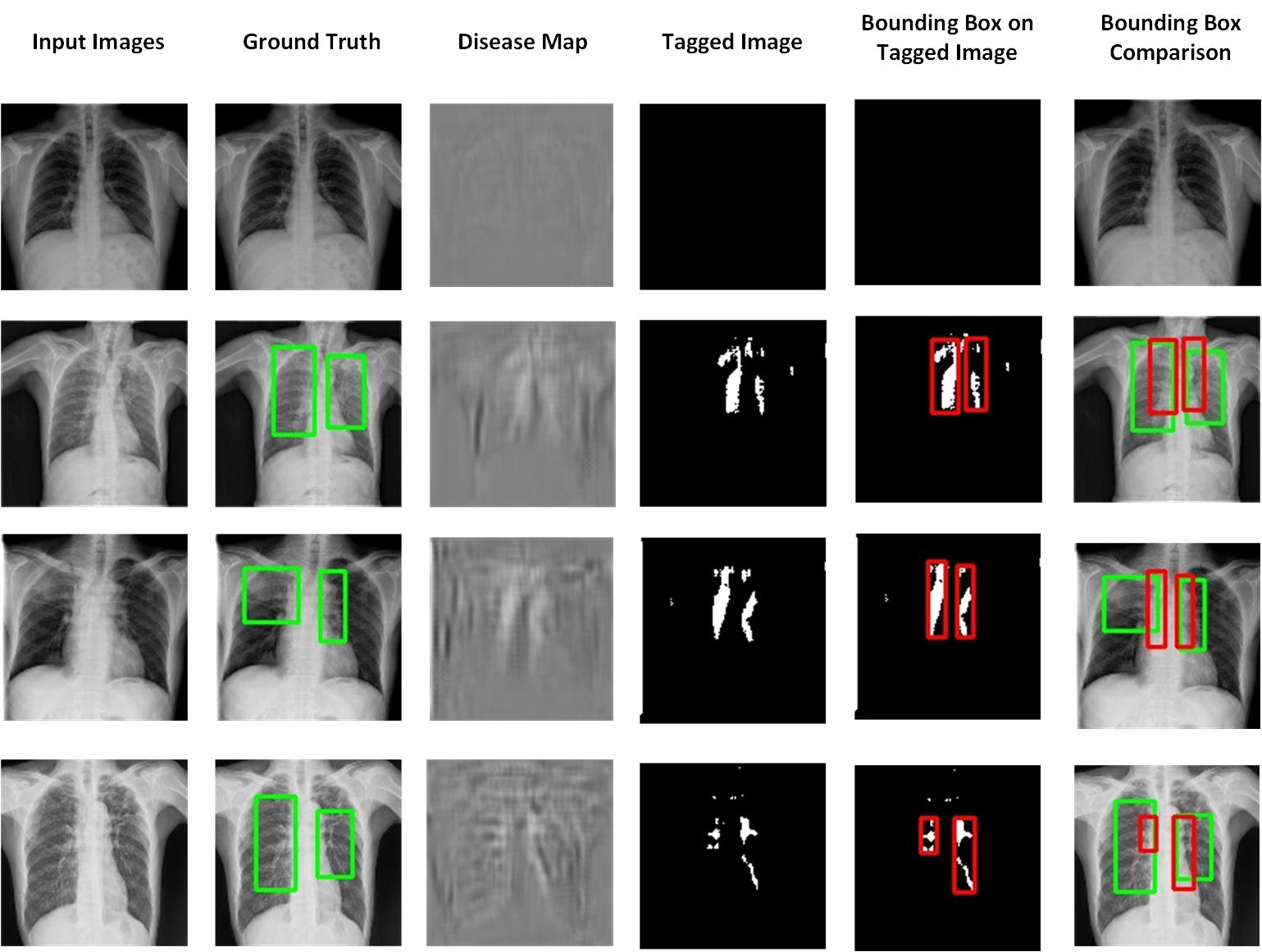}
		\caption{TBX11K with equal bounding boxes}
		\label{fig:Picture3}	
	\end{figure}
    
\subsection{Fine-grained Bounding Boxes are Important}

To streamline evaluation, bounding boxes were initially drawn on generated binary masks in numbers equal to ground truth bounding boxes. While this simplifies the quantification, it compromises the precise assessment of performance by including true negatives. To address this limitation, bounding boxes were drawn only around true positive pixels to focus exclusively on the algorithm's correct detections. The last three columns of Figure \ref{fig:Picture4} are refined evaluation approach to visualize the algorithm' identified true positive pixels and show bounding boxes drawn only around those pixels. The final comparative column is particularly noteworthy, demonstrating its capability to achieve accurate pixel-level disease localization.

 \begin{figure}[h!]
		\centering
		\includegraphics[width=1\linewidth, height=11cm]{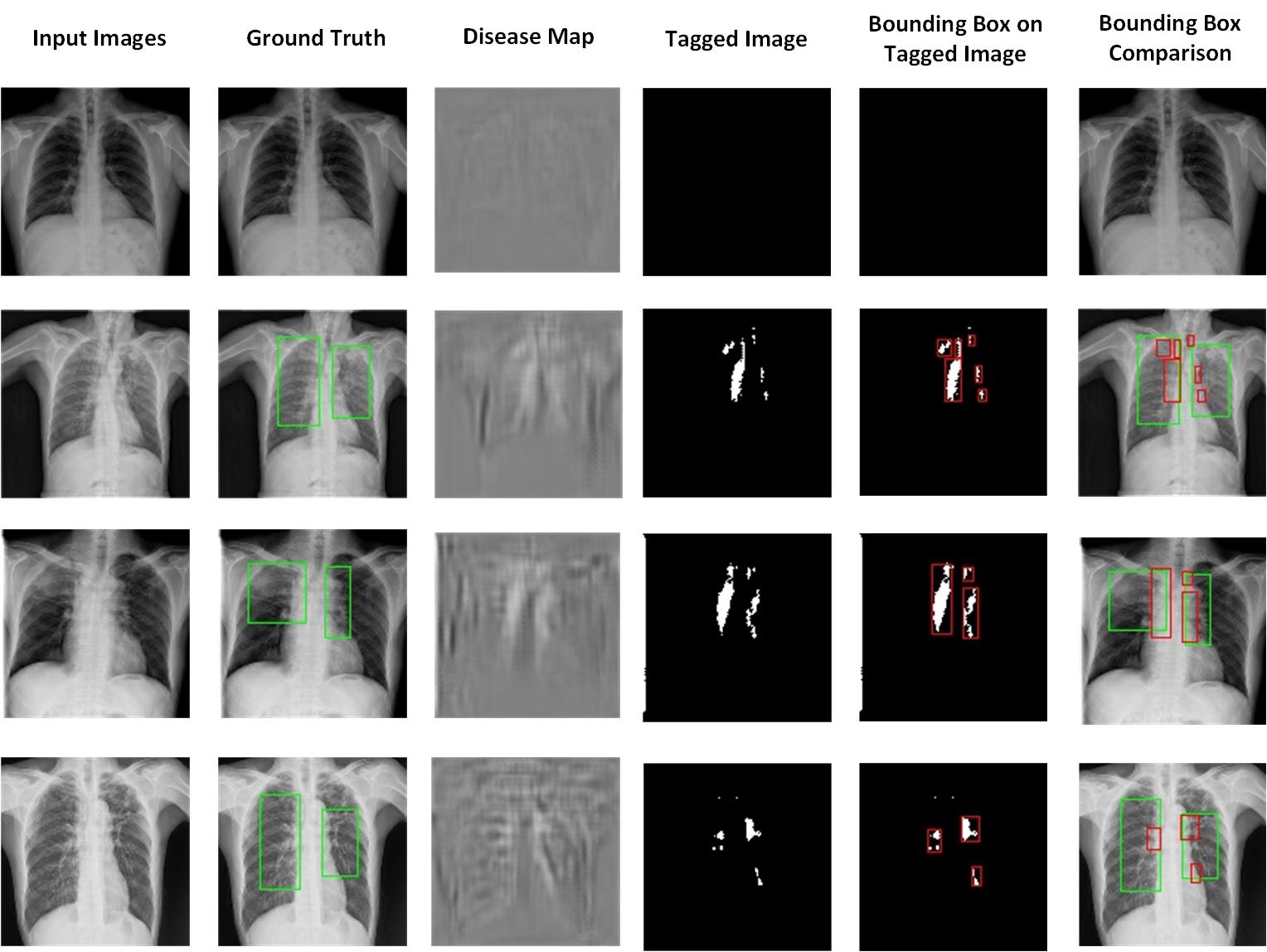}
		\caption{TBX11K with unequal bounding boxes}
		\label{fig:Picture4}	
	\end{figure}

\subsection{Only True-Positive Pixels Matter}
Figure \ref{fig:Picture5} compares results on three algorithms on TBX11K. The proposed algorithm, results shown in the second and third columns, achieves precise pixel-level disease localization and tagging. The VA-GAN model outputs with noise and limited interpretability, shown in the fourth and fifth columns. This method fails to generate a proper counterfactual normal image from the input abnormal image, often introducing random noise into the disease map during the translation process. As a single-cycle Wasserstein-GAN approach, VA-GAN lacks identity preservation, leading to the conversion of true positive pixels into false-positives by altering pixel intensities during abnormal-to-normal transformations. The final column presents CAM-based saliency maps, which visualize attention on all areas focused during decision process, wrongly classifying true-negatives as true-positive. This highlights the proposed method’s superiority in identifying true-positive pixels at a fine-grained level.

 \begin{figure}[!ht]
		\centering
		\includegraphics[width=1\linewidth, height=10cm]{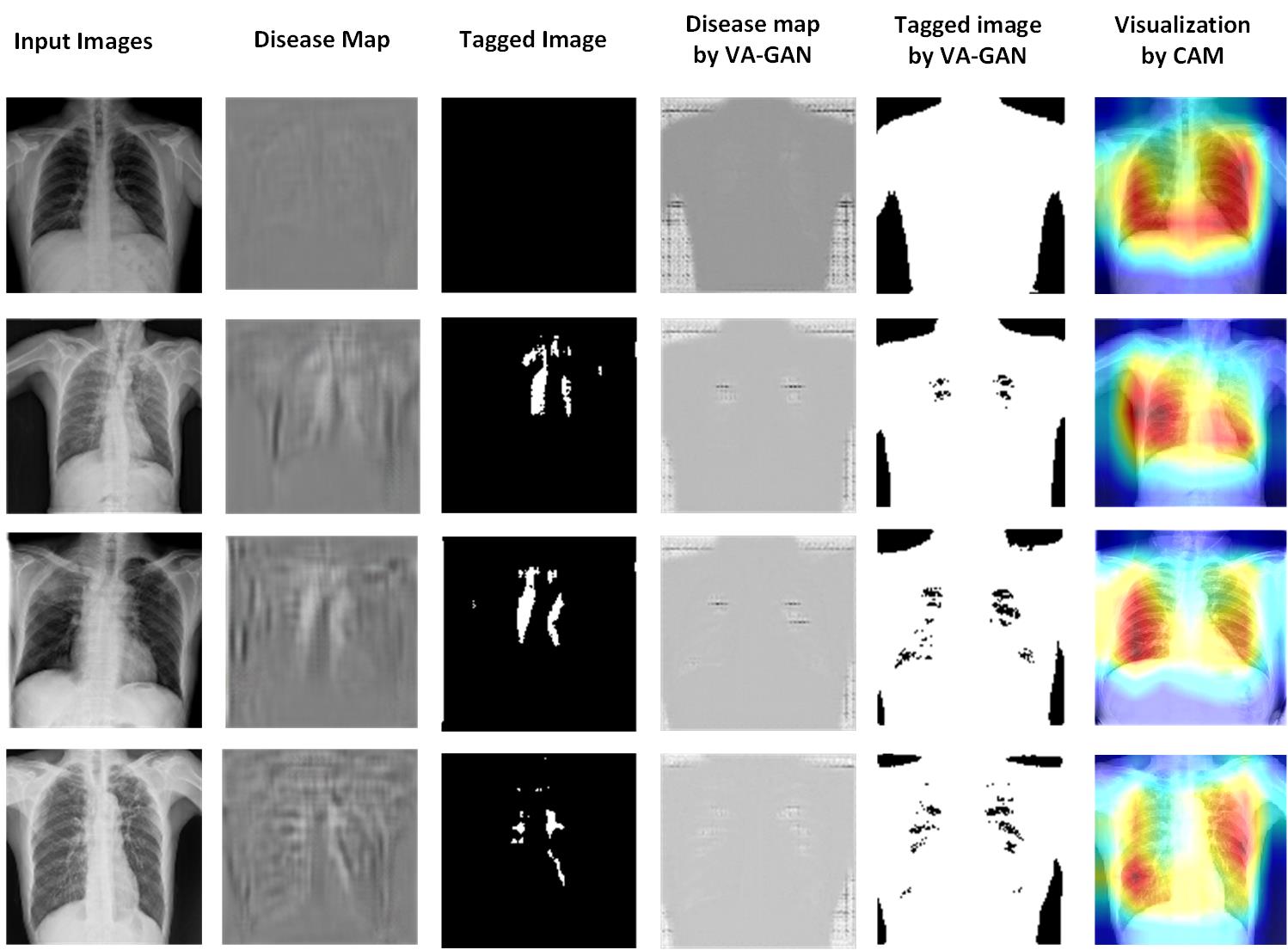}
		\caption{Comparative results}
		\label{fig:Picture5}	
	\end{figure}

\subsection{Multi-Domain Data Analysis on CheXpert}
The CheXpert dataset, comprising 14 classes with image-level annotations, was used to evaluate the proposed model’s performance in a multi-domain medical imaging context. Due to the absence of bounding boxes and pixel-level ground truth, we collaborated with radiologists to manually create bounding boxes for a subset of the samples. These labeled samples, shown in the column titled "Ground Truth," were used as a reference for model evaluation. The proposed algorithm, trained on the CheXpert dataset, generates domain-specific disease maps that show fine-grained pixel-level visualizations of diseased lesions shown in Figures \ref{fig:Picture6}  and \ref{fig:Picture7}.
These binary masks, shown in the fourth and fifth columns, enable a direct comparison with the manually created bounding boxes by radiologists, which serve as the ground truth for domain-specific diseased pixels. The complexity of this task lies in the absence of binary masks during training and the multi-class nature of the dataset. The novelty of algorithm is to learn and identify disease-specific pixels directly from image-level labels and input images. The proposed model effectively generates fine-grained and domain-specific disease maps as shown in second last column.
This is particularly significant because a single radiologist or domain expert often struggles to identify and tag different types of diseases precisely. Our algorithm is a unified framework that handles multi-disease  fine-grained visualizations and precise identification of diseased pixels. 

\begin{figure}[!ht]
		\centering
		\includegraphics[width=0.97\linewidth, height=12cm]{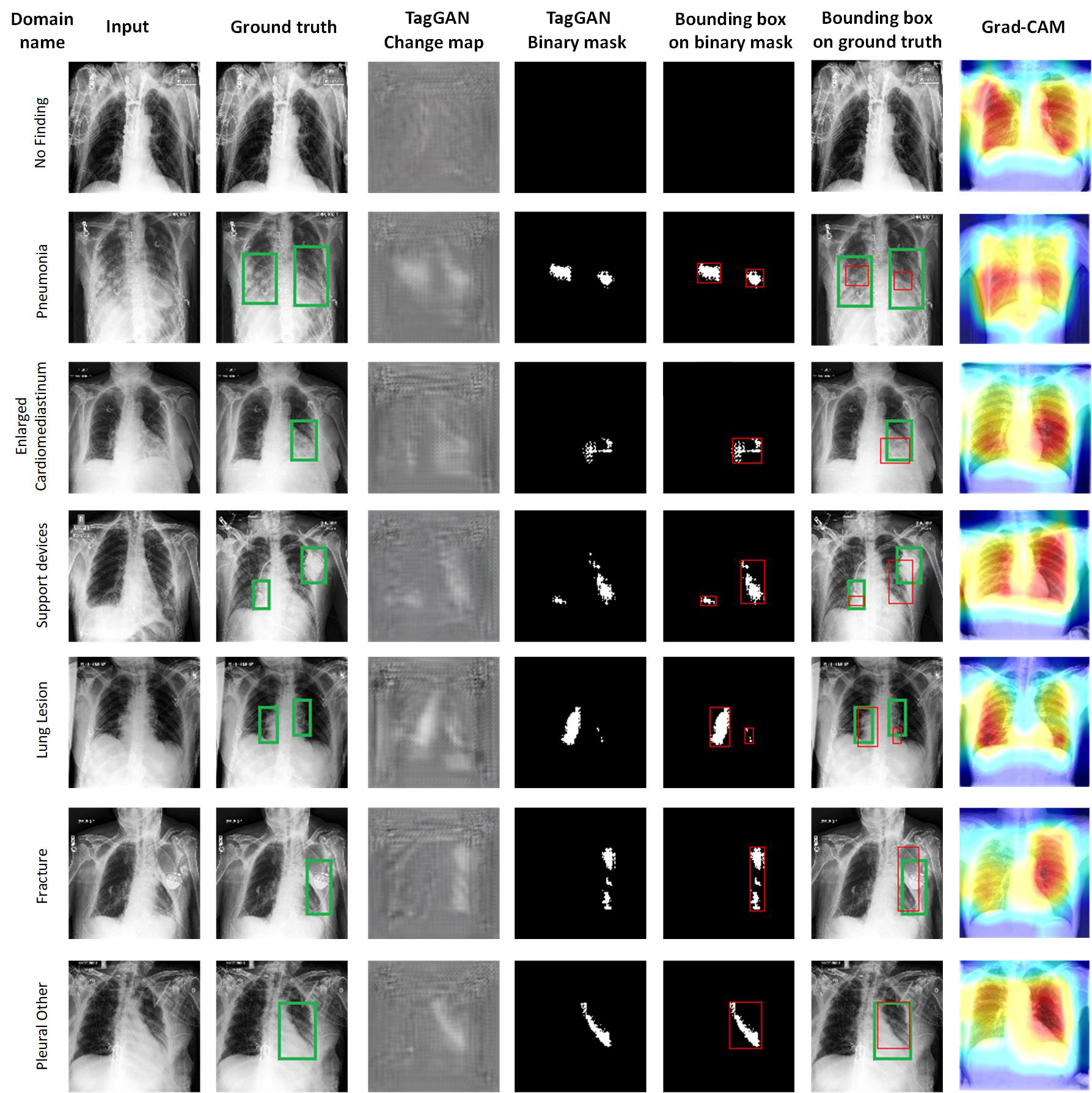}
		\caption{First seven classes of CheXpert}
		\label{fig:Picture6}	
	\end{figure}
 
 Figure \ref{fig:Picture9} compares the proposed model with existing methods where third column displays radiologist-annotated bounding boxes as ground truth. The fourth and fifth columns show bounding boxes generated by TagGAN and VA-GAN models, respectively. The sixth column overlays these bounding boxes on the ground truth for direct comparison, proving the proposed model’s precision in identifying pixel-level disease features. Comparative analysis underscores the precise alignment of our model's outputs with the ground truth.

  \begin{figure}[!ht]
		\centering
		\includegraphics[width=0.95\linewidth, height=12cm]{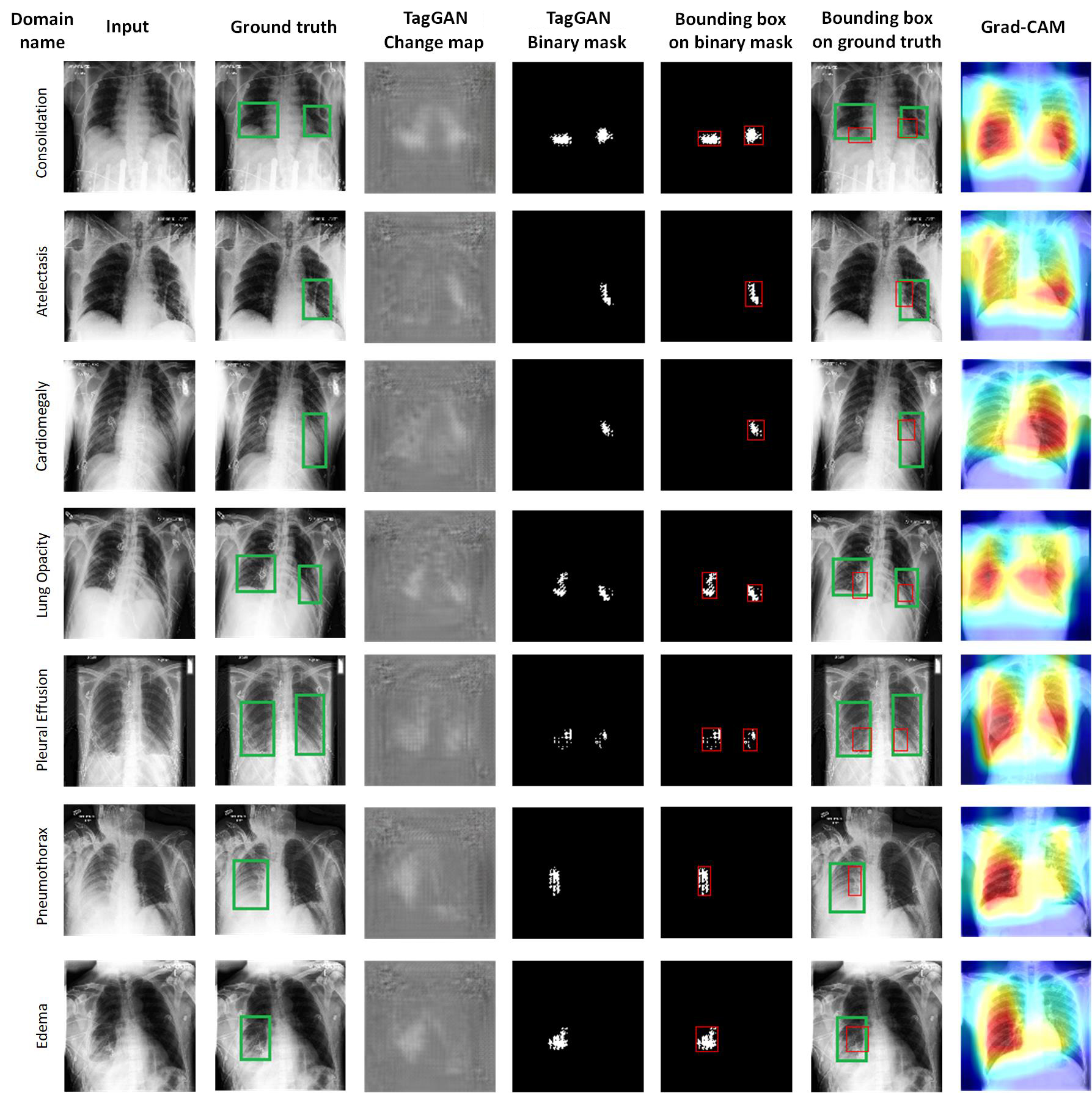}
		\caption{Second seven classes of CheXpert}
		\label{fig:Picture7}	
	\end{figure}

 \subsection{True-Negative Pixels: A Precise Take}

Figure \ref{fig:Picture8} provides an analysis of the model's behavior in cases where input image falls under 'Healthy' category. The generated disease map and its corresponding visualization for healthy image show an absence of disease-specific features, which is accurately reflected in the blank binary mask, counterfactually classifying the input as healthy.
This analysis proves the algorithm's robustness in handling true-negative cases by precisely excluding non-disease-specific regions in both the disease map visualization and binary mask generation. 
 \begin{figure}[!htb]
		\centering
		\includegraphics[width=0.8\linewidth, height=2.6cm]{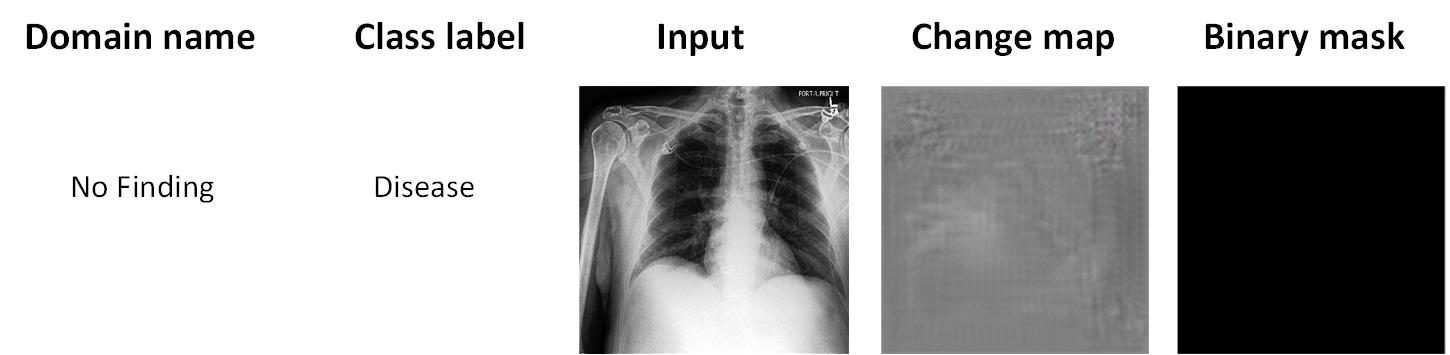}
		\caption{Normal image with disease label}
		\label{fig:Picture8}	
	\end{figure}
 
 \subsection{Qualitative Results on COVID-19}
Figure \ref{fig:Picture10} is structured to present COVID-19 results sequentially from input images to tagged outputs. The Input Images column contains three COVID-diseased X-rays and one normal X-ray used for evaluation. The third column, labeled Disease Map, highlights disease-specific pixels while suppressing irrelevant areas. For normal X-rays, the absence of disease is reflected in a blank binary mask, demonstrating the model's specificity. Disease maps provide pixel-specific visualization of COVID-19 by sharpening affected areas effectively.

 \begin{figure}[!ht]
		\centering
		\includegraphics[width=0.97\linewidth, height=13cm]{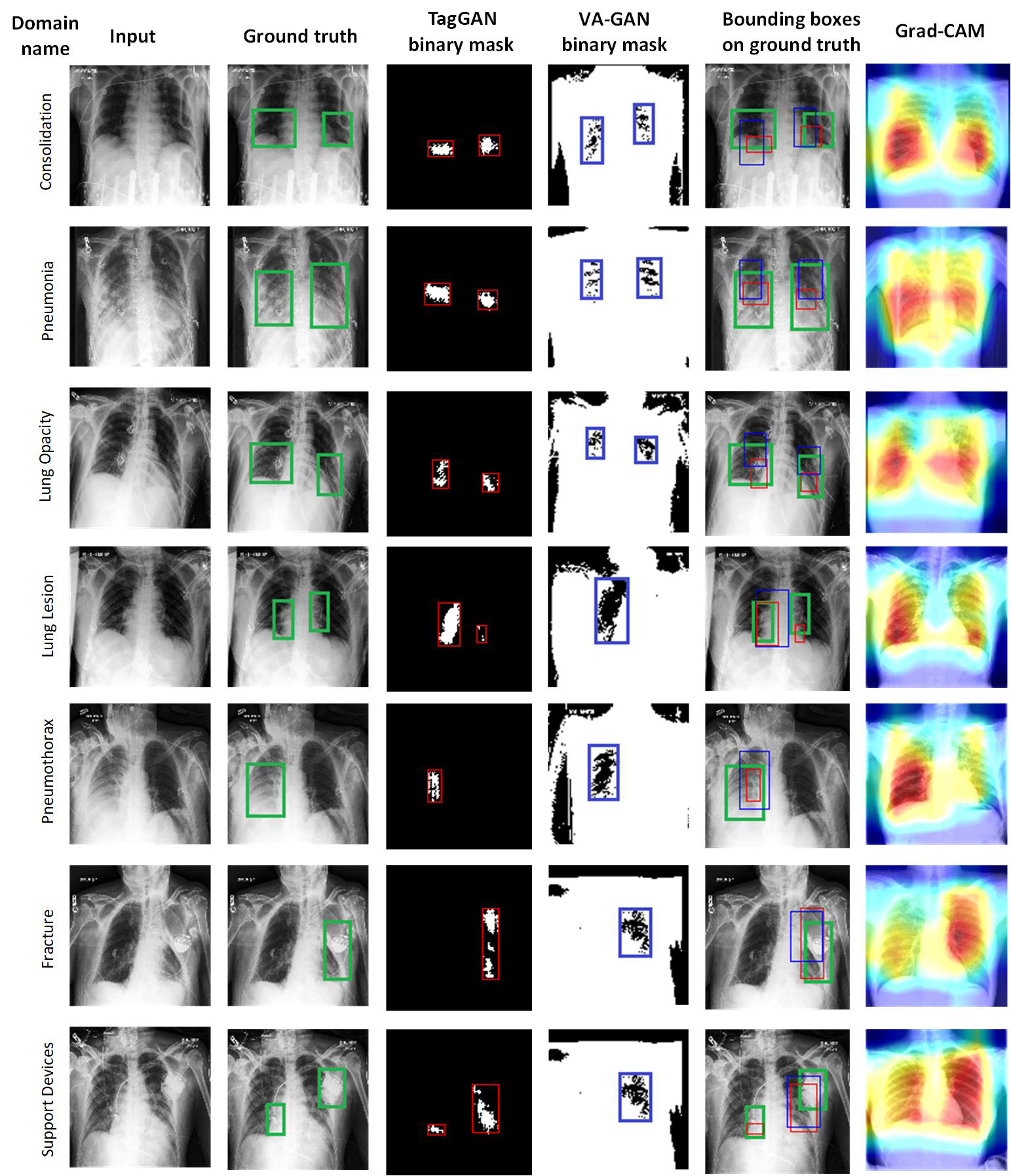}
		\caption{Comparison on CheXpert}
		\label{fig:Picture9}	
	\end{figure}

  \begin{figure}[ht!]
		\centering
		\includegraphics[width=0.5\linewidth, height=9cm]{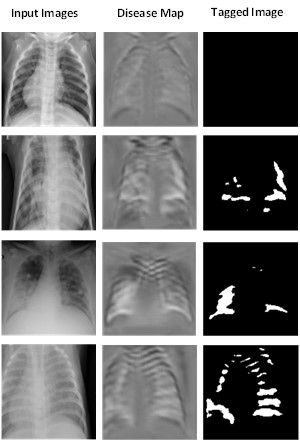}
		\caption{COVID-19 results}
		\label{fig:Picture10}	
	\end{figure}
 }

\section{Conclusion}

This study presents TagGAN, a novel generative adversarial network-based framework for data tagging in weakly supervised settings, specifically addressing the challenge of generating pixel-level annotations from image-level labeled chest radiographs. By identifying and visualizing true-positive pixels on disease maps and generating binary mask annotations, TagGAN overcomes the limitations of traditional methods that rely on pixel-level ground truth, making it particularly relevant for clinical applications.

Unlike prior discrimination-based approaches, which are constrained by their dependence on detailed annotations, TagGAN demonstrates the ability to preserve key anatomical details during the generation of normal-abnormal image pairs, ensuring both robustness and interpretability. The model's ability to operate effectively without pixel-level annotations significantly enhances its utility in real-world medical imaging scenarios, where such annotations are often unavailable. 

This work addresses a critical gap in medical imaging and  lays the foundation for future research in explainable AI for healthcare. By reducing the reliance on manual annotation and providing precise disease visualization, TagGAN has the potential to streamline workflows for radiologists and improve diagnostic accuracy. We anticipate that this approach will inspire further advancements, encouraging the development of more accurate, scalable, and interpretable systems to support the healthcare sector.

\section{Acknowledgments}
This work was supported by International Research Scholarship by University of Technology Sydney, Australia.

\section{Ethics Statement}
This study exclusively utilized publicly available datasets and did not involve the collection or analysis of private, sensitive, or identifiable human data. No human participants, animals, or clinical trials were involved in this research. As such, no additional ethical approval was required.

\end{document}